\documentclass[preprint, 11pt, authoryear]{elsarticle}

\usepackage[T1]{fontenc}
\usepackage{microtype}
\usepackage{geometry}
\usepackage{booktabs}
\usepackage{tabularx}
\newcolumntype{Y}{>{\centering\arraybackslash}X}
\usepackage{makecell}
\newcommand*{\tabindent}{\hspace{3mm}}
\usepackage{caption}
\usepackage{bm}
\usepackage[hyphens]{url}
\usepackage[hidelinks]{hyperref}
\hypersetup{breaklinks=true}
\journal{Social Science \& Medicine}

\begin{document}

\begin{frontmatter}

\title{Glucose Control, Sleep, Obesity, and Real-World Driver Safety at Stop Intersections in Type 1 Diabetes}

\author[inst1]{Ashirwad Barnwal\corref{cor1}}
\ead{ashirwad@iastate.edu}
\cortext[cor1]{Corresponding author}
\author[inst1]{Anuj Sharma}
\author[inst2]{Luis Riera-Garcia}
\author[inst1]{Koray Ozcan}
\author[inst1]{Sayedomidreza Davami}
\author[inst2]{Soumik Sarkar}
\author[inst3]{Cyrus Desouza}
\author[inst4]{Matthew Rizzo}
\author[inst4]{Jennifer Merickel}

\affiliation[inst1]{
    organization = {Institute for Transportation (InTrans), Iowa State University},
    city = {Ames},
    postcode = {50010},
    state = {Iowa},
    country = {United States}
}
\affiliation[inst2]{
    organization = {Department of Mechanical Engineering, Iowa State University},
    city = {Ames},
    postcode = {50011}, 
    state = {Iowa},
    country = {United States}
}
\affiliation[inst3]{
    organization = {Division of Diabetes, Endocrinology, and Metabolism, University of Nebraska Medical Center},
    city = {Omaha},
    postcode = {68198},
    state = {Nebraska},
    country = {United States}
}
\affiliation[inst4]{
    organization = {Department of Neurological Sciences, University of Nebraska Medical Center},
    city = {Omaha},
    postcode = {68198}, 
    state = {Nebraska},
    country = {United States}
}

\begin{abstract}
\textbf{Background}: Diabetes is associated with obesity, poor glucose control and sleep dysfunction which impair cognitive and psychomotor functions, and, in turn, increase driver risk. How this risk plays out in the real-world driving settings is terra incognita. Addressing this knowledge gap requires comprehensive observations of diabetes driver behavior and physiology in challenging settings where crashes are more likely to occur, such as stop-controlled traffic intersections, as in the current study of drivers with Type 1 Diabetes (T1DM). \textbf{Methods}: 32 legally licensed active drivers (18 T1DM + 14 controls) from around Omaha, NE participated in 4-week, real-world study. Each participant's own vehicle was instrumented with an advanced telematics and camera system collecting driving sensor data (e.g., speed, acceleration, GPS) and video (forward road and cabin). Videos were analyzed using computer vision models detecting traffic elements to identify stop signs. Stop sign detections and driver stopping trajectories were clustered to geolocate and extract driver-visited stop intersections. Driver videos were then annotated to record stopping behavior (full stop, rolling stop, no stop) and key traffic characteristics (e.g., presence/absence of lead/crossing vehicles). Stops were categorized as safe or unsafe based on traffic law. \textbf{Results}: Mixed effects logistic regression models examined how stopping behavior (safe vs. unsafe) in T1DM drivers was affected by 1) abnormal sleep (above or below 7--9 hours), 2) obesity, and 3) poor glucose control (greater standard deviation [SD], coefficient of variation [CV], low blood glucose index [LBGI], and high blood glucose index [HBGI]). Model results indicate that one standard deviation increase in BMI (\(\sim\)7 points) in T1DM drivers associated with a 14.96 increase in unsafe stopping odds compared to similar controls. Abnormal sleep and glucose control were not associated with increased unsafe stopping. \textbf{Conclusion}: This study links chronic patterns of abnormal T1DM driver physiology, sleep, and health to driver safety risk at intersections, advancing models to identify real-world safety risk in diabetes drivers for clinical intervention and development of in-vehicle safety assistance technology.
\end{abstract}

\begin{keyword}
naturalistic driving \sep unsafe stopping \sep driver risk \sep type 1 diabetes \sep sleep dysfunction \sep obesity \sep poor glucose control
\end{keyword}

\end{frontmatter}

\section{Introduction}
Our goal is to develop data-driven assessments of diabetes patient risk from real-world safety behavior that alert patient clinical care teams to worsening disease or glucose control. To meet this goal, we quantified remote, digital health profiles in drivers with type 1 diabetes (T1DM) that link glucose control, obesity, and sleep dysfunction to geo-specific driving risk. T1DM is a multi-factorial disease affecting regulation and metabolism of blood glucose \citep{kahanovitz_type_2017}. In 2018, nearly 1.4 million adults (age \(\geq\) 20 years) had T1DM among >26 million adults in the United States (US) with diabetes generally \citep{centers_for_disease_control_and_prevention_national_2020}. Diabetes prevalence is increasing and projections estimate 4.4 million US adults will have T1DM by 2050 \citep{dabelea_prevalence_2014}.

Diabetes is associated with cognitive, autonomic, and psychomotor impairments \citep{zilliox_diabetes_2016} that increase driving risk \citep{merickel_at-risk_2017}. Driver impairments arise from poor glucose control indexed by glucose variability and overall glucose level (low: hypoglycemia; high: hyperglycemia). Poor glucose control also disrupts sleep and increases risk of obesity, which can independently worsen glucose control and produce impairment \citep{larcher_sleep_2015}.

\subsection{Sleep}
Excessive daytime sleepiness (EDS), often resulting from sleep disruption \citep{slater_excessive_2012}, affects over 20\% of the population \citep{ohayon_wakefulness_2008, ohayon_epidemiological_2011} and can worsen health outcomes, cognitive function, driver safety, and quality of life. Sleep disruption and EDS affect executive \citep{regestein_self-reported_2004}, memory \citep{sternberg_largest_2013}, and attention \citep{krieg_neurobehavioral_2001} functions, impacting daily activities like safe vehicle operation. EDS is a hallmark feature of diabetes. In diabetes, sleep dysfunction, EDS, and resulting cognitive impairment can result from nighttime hypoglycemia (prevalent among T1DM patients). Poor sleep can feedback to worsen glucose control, increasing risk.

Beyond diabetes, EDS and driver fatigue are critical problems of public health and safety. EDS was implicated in 72,000 vehicle crashes and 44,000 crash injuries in 2013 \citep{connor_role_2001, drake_10-year_2010, stutts_driver_2003}. The American Academy of Sleep Medicine and Sleep Research Society recommends 7--9 hours of nightly sleep for the typical adult \citep{watson_recommended_2015}. Despite that, \(\sim\)35\% of American adults (18--60 years) sleep <7 hours nightly \citep{liu_prevalence_2016}. Co-morbid health conditions common in diabetes, like obstructive sleep apnea, also worsen sleep and cause daytime impairment \citep{peppard_increased_2013}.

Sleep can fluctuate widely over months, days, and even hours \citep{merickel_driving_2019, tippin_sleep_2016} and can be indexed with actigraphy \citep{cook_optimizing_2019, filardi_actigraphic_2015}. The rapid growth in the use of wearable technology over the past several years has made actigraphy devices like smartwatches a popular method for tracking real-world sleep patterns. Actigraphy-derived sleep data has been validated to have 80\%, on average, correspondence to polysomnography data across multiple nights and populations \citep{marino_measuring_2013}. This motivates the use of actigraphy-based sleep measurements in our study.

\subsection{Obesity}
Obesity is a growing public health concern and a leading risk factor for diabetes \citep{algoblan_mechanism_2014, ayton_obesity_2019}. Obesity is defined as having a body mass index (BMI: ratio of weight in kilograms to height in meters squared) of \(\geq\)30 \citep{hales_prevalence_2020}. Generally, obesity has increased in the population due to genetic and lifestyle factors (e.g., sedentary behavior, poor dietary habits) \citep{chaput_findings_2014}. Obesity increases the risk of poor glucose control, particularly hyperglycemia, and co-morbid health conditions like neuropathy, retinopathy, and cardiovascular disease---all of which can worsen cognitive function and glucose control. Health conditions that produce sleep disruption, like obstructive sleep apnea, also associate with obesity. These factors link obesity to driver safety risk.

\subsection{Glycemic variability}
Glycemic variability (GV) indexes a patient's degree of glucose control by quantifying how much glucose levels vary between high (\(\geq\)180 mg/dL) and low (\(\leq\)70 mg/dL), providing more detail on glucose fluctuations over time than average Hemoglobin A1c (HbA1c) \citep{devries_glucose_2013, nathan_relationship_2007}. Poor glucose control (high or low fluctuations) is associated with acute and persistent cognitive impairment, sleep disruption and EDS, obesity, and worsening health in diabetes---increasing driver safety risk \citep{carr_older_2010, huisingh_general_2018, kim_association_2015}. GV can be indexed across multiple measures, each indexing different aspects of impaired glucose control. For this paper, we compared the impact of clinically-validated metrics of glucose control on driver risk overall (standard deviation [SD], coefficient of variation [CV]), due to hypoglycemia (low blood glucose index [LBGI]), and due to hyperglycemia (high blood glucose index [HBGI]) \citep{kovatchev_glucose_2016}:

\begin{itemize}
\item
  SD quantifies the overall variability in glucose around the mean.
\item
  CV quantifies the magnitude of variability in glucose relative to mean.
\item
  LBGI quantifies the extent and frequency of hypoglycemic excursions.
\item
  HBGI quantifies the extent and frequency of hyperglycemic excursions.
\end{itemize}

\subsection{Highway focus area}
Stop sign-controlled intersections (hereafter referred to as simply ``stop intersections'') are a type of roadway intersection where at least one of the approaches is controlled by a stop sign. Drivers are required by law to come to a complete stop at these intersections. Failure to stop (e.g., rolling stop, no stop) can result in a traffic citation. Intersections generally are critical safety locations and are collectively responsible for over 33\% of police-reported motor vehicle crashes in the US every year \citep{choi_crash_2010, cunard_signalized_2003}. Across intersection types, stop intersections have the highest crash frequency and severity \citep{campbell_analysis_2004, hauer_safety_1997}. NHTSA indicates that annually about 700,000 traffic crashes in the US occur at stop intersections, with nearly one-third resulting in injuries \citep{retting_analysis_2003}. Another study concluded that nearly 70\% of all crashes that occurred at stop intersections resulted from stop sign violations \citep{retting_analysis_2003}. Driver impairments, like diabetes-related cognitive and psychomotor dysfunctions from sleep, glucose, and obesity impairments, may increase risk of unsafe stopping at intersections \citep{choi_crash_2010}. These factors make stop intersections a key node to investigate the role of sleep, obesity, and GV in T1DM driver safety. While several factors affect driver safety, for the purpose of this analysis we defined ``safe'' stops as those following traffic law (a complete stop before entering the intersection).

\section{Hypothesis}
We tested the hypothesis that unsafe stopping behavior in T1DM drivers would be associated with 1) abnormal sleep (above or below 7--9 hours), 2) greater BMI, and 3) poor glucose control indexed by GV (greater SD, CV, LBGI, and HBGI). Sleep and obesity were assessed using data from T1DM and control participants without diabetes. Glucose control was assessed using data only from T1DM participants.

\section{Methods}

\subsection{Data sources}

\subsubsection{Participants}
A total of 36 active, legally licensed drivers were recruited at the University of Nebraska Medical Center (UNMC) via community or clinic recruitment. Participants were from the Omaha, Nebraska community and the surrounding areas (age range: 21--55 years; mean = 32.2 years). Out of 36 participants, 20 (56\%) had T1DM and 16 (44\%) were controls without diabetes. Controls were matched to diabetes drivers across age (\(\pm\) 6 years), gender, education (\(\pm\) 2 years), and driving season (winter vs. not winter). Control participants provided baseline data on driver behavior in the absence of diabetes. Participants consented to the study following institutional guidelines (UNMC IRB \#462-16-FB).

\begin{figure}[!htbp]
\centering
\includegraphics[width = \textwidth]{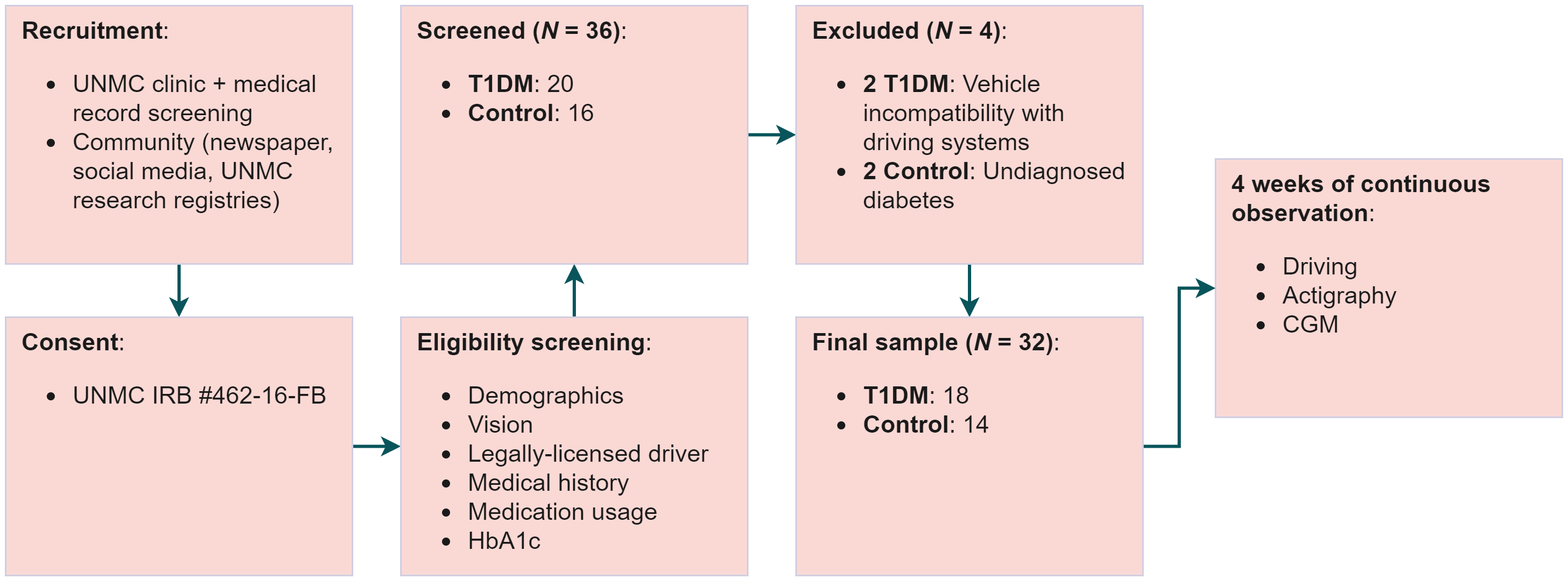}
\caption{Flow diagram of study participation}
\label{fig:study-flow-diag}
\end{figure}

Eligibility was assessed at study start via self-reported demographics, medical exam by an endocrinologist, self-reported medical history and medication usage, and HbA1c blood labs. T1DM participants had HbA1c \(\geq\) 6.5\% and a diagnosis of type 1 diabetes. Controls did not have diabetes and had HbA1c <5.7\% \citep{american_diabetes_association_standards_2019}. All participants had (a) safe vision for driving (corrected or uncorrected visual acuity <20/50 OU) per Nebraska licensure regulations; (b) absence of significant, confounding medical conditions (peripheral nerve, eye, renal, neurological, and major psychiatric diseases); and (c) no use of confounding medications (narcotics, sedating antihistamines, and major psychoactive medication). Four participants were excluded based on eligibility screening: 1) two T1DM participants were excluded due to vehicle incompatibility with the study driving systems, and 2) two controls were excluded due to undiagnosed diabetes based on HbA1c levels. The final sample included 32 participants (T1DM: \emph{N} = 18; Controls: \emph{N} = 14). See Figure \ref{fig:study-flow-diag} for the participant flow diagram.

\subsubsection{Study design}
\textbf{Black Box}: Each participant's primary vehicle was instrumented with an advanced, custom telematics system called a Black Box to collect driving data for 4 weeks. The Black Box system contained two cameras (forward roadway, cabin), a cabin microphone, GPS, accelerometer, and an on-board diagnostics (OBD) port reader to collect vehicle data like speed and engine information. The system was affixed on the windshield behind rearview mirror and did not obstruct the driver's roadway view. The Black Box system collected data at a frequency of 1 Hz from on ignition to off ignition.

\textbf{In-Laboratory Assessments}: At study start, data on demographics were collected for each participant using a standard questionnaire that assessed their age, gender, race/ethnicity, marital status, and socioeconomic status. Medical history and medication usage were collected via self-report using a standardized intake. Data on visual acuity [Early Treatment Diabetic Retinopathy Study (ETDRS) OU] and contrast sensitivity [ETDRS, 2.5\% OU] were also collected. A descriptive summary of the key demographic and medical characteristics of study participants is presented in Table \ref{tab:part-smry-table}.

\begin{table}[!htbp]
\centering
\caption{Summary of demographic and medical characteristics for study participants}
\label{tab:part-smry-table}
\begin{tabular}{@{}lccc@{}}
\toprule
 & \textbf{T1DM (\textit{N} = 18)} & \textbf{Control (\textit{N} = 14)} & \textbf{Total (\textit{N} = 32)} \\ \midrule
\textbf{Age (years)} &  &  &  \\
\tabindent Mean (SD) & 31.22 (9.67) & 33.36 (10.45) & 32.16 (9.91) \\
\tabindent Range & 21.00--52.00 & 21.00--55.00 & 21.00--55.00 \\
\textbf{Gender} &  &  &  \\
\tabindent Female & 11 (61.1\%) & 10 (71.4\%) & 21 (65.6\%) \\
\tabindent Male & 7 (38.9\%) & 4 (28.6\%) & 11 (34.4\%) \\
\textbf{Race} &  &  &  \\
\tabindent Asian & 0 (0.0\%) & 1 (7.1\%) & 1 (3.1\%) \\
\tabindent White & 18 (100.0\%) & 13 (92.9\%) & 31 (96.9\%) \\
\textbf{Driving experience (years)} &  &  &  \\
\tabindent Mean (SD) & 15.61 (9.59) & 16.07 (9.86) & 15.81 (9.55) \\
\tabindent Range & 6.00--36.00 & 4.00--33.00 & 4.00--36.00 \\
\textbf{HbA1c (\%)} &  &  &  \\
\tabindent Mean (SD) & 7.79 (1.05) & 5.19 (0.30) & 6.65 (1.54) \\
\tabindent Range & 6.80--11.30 & 4.70--5.70 & 4.70--11.30 \\ \bottomrule
\end{tabular}
\end{table}

\textbf{CGM}: T1DM participants in this study were provided with a Food and Drug Administration (FDA) approved Dexcom G4 PLATINUM Professional Continuous Glucose Monitor (CGM). Participants were trained on CGM use at study start. The CGM was worn at the waist for the 4-weeks of data collection and measured glucose readings from interstitial fluid via a small wire underneath the skin. Glucose readings were taken at five-minute intervals. CGM data was blinded from participants. Glucose readings were processed using FDA guidelines for CGM data quality \citep{center_for_devices_and_radiological_health_fda_2014}, details of which are available in \cite{barnwal_sugar_2021}. The processed glucose readings were used to compute \textbf{GV metrics SD, CV, HBGI, and LGBI}.

\textbf{Sleep}: Study participants wore Fitbit Blaze actigraphy monitor on their non-dominant wrist for the 4-week study period. The device collected data on sleep duration and quality. Wrist worn Fitbits and other actigraphy devices have been validated against polysomnography for accurate sleep and wakefulness detection \citep{haghayegh_accuracy_2019}. Participants did not receive feedback on sleep throughout the study. Nightly sleep was averaged across the study period. Sleep duration was assessed continuously from less to more sleep, on average.

\textbf{Obesity}: Participant height and weight data was collected during medical exam at study start. BMI was computed from height and weight. Obesity was assessed from lesser to greater obesity.

\subsubsection{Stop behavior extraction}
\begin{figure}[!htbp]
\centering
\includegraphics{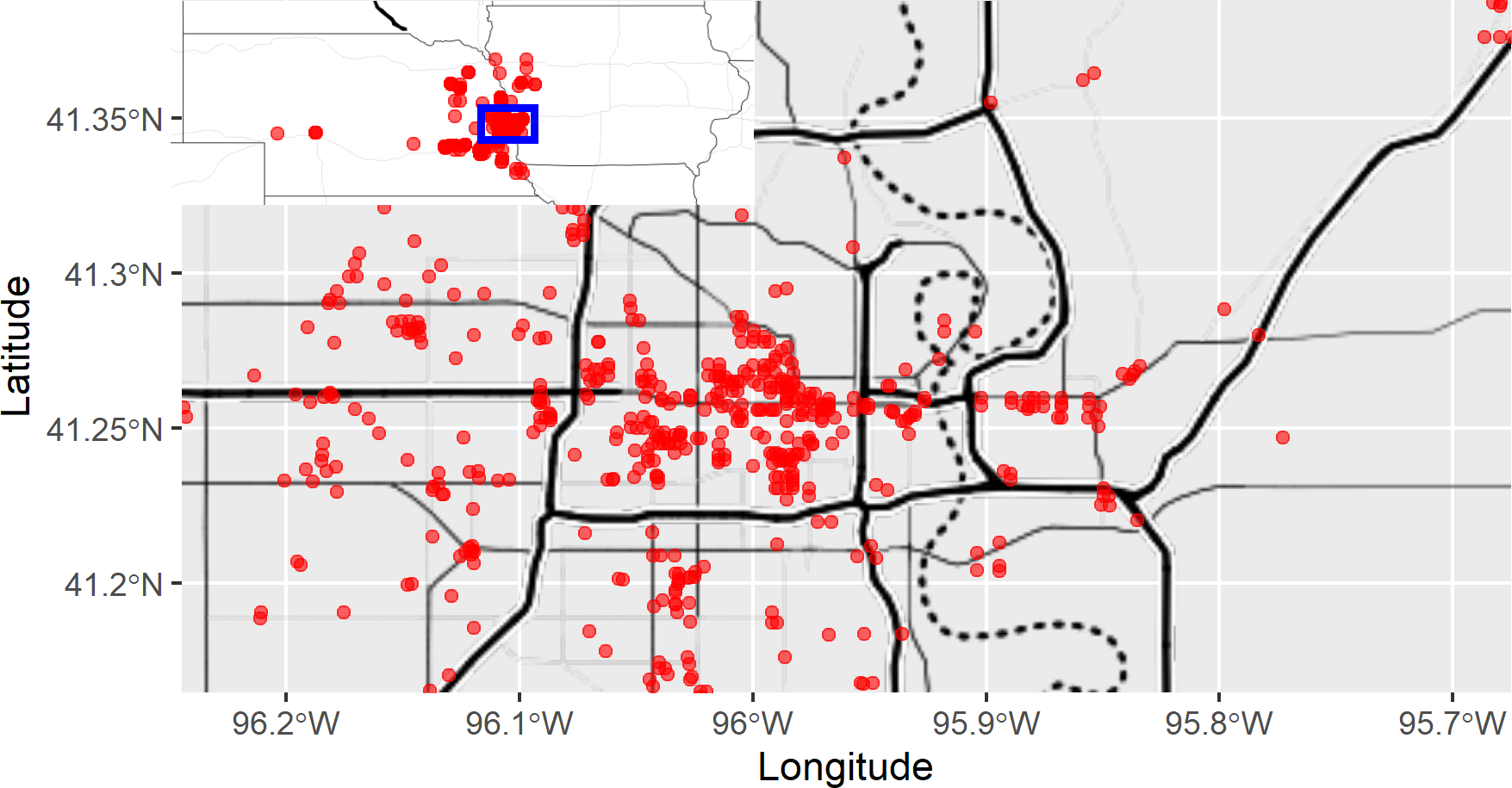}
\caption{Geographic locations of participant-visited stop intersections. The blue rectangle in the figure highlights the areas of Omaha, Nebraska and Council Bluffs, Iowa, which were this study's primary recruitment area and had the highest density of stop intersections visited by the participants.}
\label{fig:stop-intxn-map}
\end{figure}

To identify the list of stop intersections visited by the participants, we used two methods: (a) For data in Nebraska, a density-based spatial clustering application with noise (DBSCAN) algorithm was applied to merged Black Box and computer vision data (stop sign detections) (see \cite{barnwal_sugar_2021, barnwal_ndsintxn_2021} for full details on the method), and (b) For data in Iowa, an Iowa stop intersection database maintained at the Institute for Transportation at Iowa State University was used. The states of Iowa and Nebraska were analyzed because the vast majority (95.5\%) of driver trips were located in these two states. The map of stop intersections visited by drivers in this study is shown in Figure \ref{fig:stop-intxn-map}.

Once the driver-visited stop intersections were identified, video clips of stopping trajectories passing through the intersections were extracted. The methods used for extracting and annotating video clips are presented in \cite{barnwal_sugar_2021, barnwal_ndsintxn_2021}. Extracted video clips were annotated by a human reviewer to record data on the following variables:

\begin{itemize}
\item
  \textbf{Stopping behavior}: A \textbf{full stop} was defined as a driver stopping at a stop intersection for at least 2 seconds (\emph{N} = 347). A \textbf{rolling stop} occurred when the driver reduced their travel speed but did not stop completely (\emph{N} = 1036). A \textbf{no stop} was coded when the driver passed through the intersection at their initial travel speed (\emph{N} = 277). Safe stops for analysis were defined as full stops. Unsafe stops were defined as rolling and no stops.
\item
  \textbf{Intersection traffic}: Traffic factors that could interrupt the driver's behavior (e.g., change in driving speed) were coded during the driver's intersection approach based on presence of 1) lead vehicle, 2) crossing vehicle, and 3) crossing pedestrian. For each traffic scenario, 3 subtypes were considered: (a) the traffic factor was absent (coded as \textbf{none present}), (b) the traffic factor was \textbf{present without effect} on the driver's behavior (e.g., the lead vehicle was sufficiently far ahead of the driver in the approach lane to not affect the driver's behavior), and (c) the traffic factor was \textbf{present with effect} if it affected the driver's behavior (e.g., participant driver was closely following a lead vehicle). Driver stop approaches containing traffic factors that were present with effect were discarded for analysis due to confound.
\item
  \textbf{Driver identity}: Information on whether a study participant or someone else was driving the vehicle during a stop sign encounter was also recorded. Data from driver's not participating in the study was discarded from analysis.
\end{itemize}

\begin{figure}[!htbp]
\centering
\includegraphics[width = \textwidth]{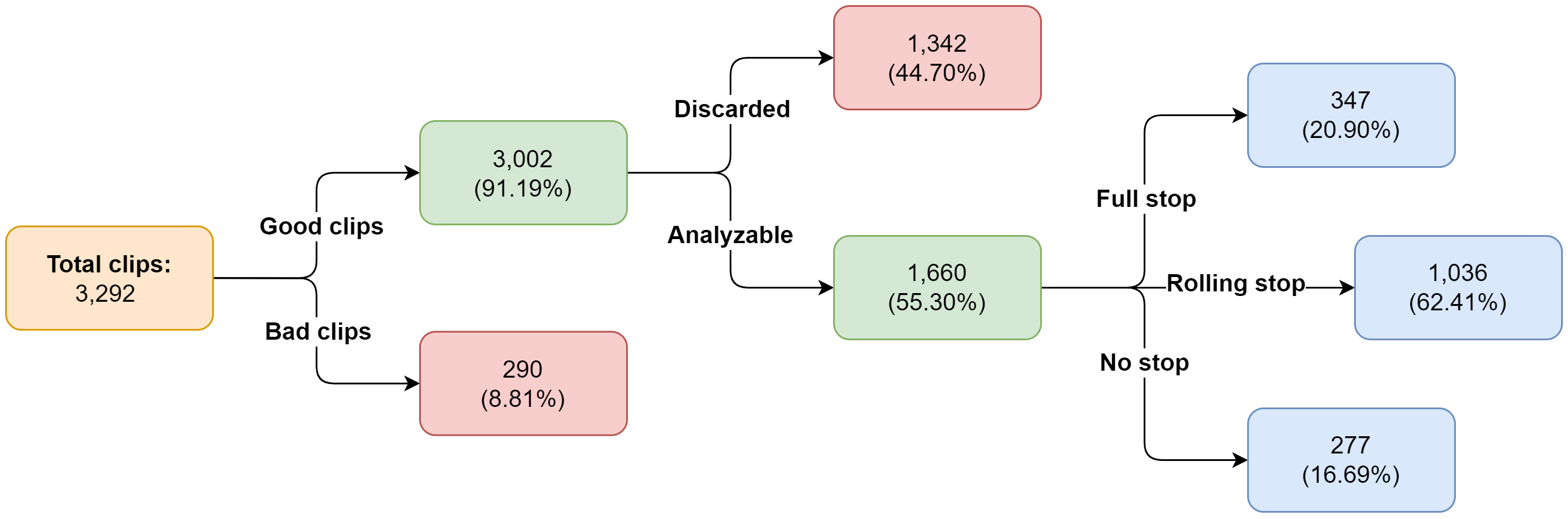}
\caption{Summary of the video review process}
\label{fig:video-review-smry}
\end{figure}

Table \ref{tab:video-vars-smry-table} provides a summary of annotated variables and Figure \ref{fig:video-review-smry} shows a high-level summary of the entire video review process.

\begin{table}[!htbp]
\centering
\small
\caption{Summary of variables extracted through manual review. ``Effect'' denotes whether or not the participant altered their vehicle speed due to the presence of this variable.}
\label{tab:video-vars-smry-table}
\begin{tabular}{@{}lccc@{}}
\toprule
 & \textbf{\makecell{Safe (Full Stop) \\ (\textit{N} = 1140)}} & \textbf{\makecell{Unsafe \\ (80.8\% Rolling, 19.2\% No Stop) \\ (\textit{N} = 1864)}} & \textbf{\makecell{Total \\ (\textit{N} = 3004)}} \\ \midrule
\textbf{\makecell[tl]{Lead vehicle \\ status}} &  &  &  \\
\tabindent None present & 861 (75.5\%) & 1500 (80.5\%) & 2361 (78.6\%) \\
\tabindent Present with effect & 214 (18.8\%) & 240 (12.9\%) & 454 (15.1\%) \\
\tabindent Present without effect & 65 (5.7\%) & 124 (6.7\%) & 189 (6.3\%) \\
\textbf{\makecell[tl]{Crossing vehicle \\ status}} &  &  &  \\
\tabindent None present & 336 (29.5\%) & 1317 (70.7\%) & 1653 (55.0\%) \\
\tabindent Present with effect & 681 (59.7\%) & 117 (6.3\%) & 798 (26.6\%) \\
\tabindent Present without effect & 123 (10.8\%) & 430 (23.1\%) & 553 (18.4\%) \\
\textbf{\makecell[tl]{Crossing pedestrian \\ status}} &  &  &  \\
\tabindent None present & 1102 (96.7\%) & 1838 (98.6\%) & 2940 (97.9\%) \\
\tabindent Present with effect & 16 (1.4\%) & 7 (0.4\%) & 23 (0.8\%) \\
\tabindent Present without effect & 22 (1.9\%) & 19 (1.0\%) & 41 (1.4\%) \\
\textbf{\makecell[tl]{Is primary participant \\ driving?}} &  &  &  \\
\tabindent No & 96 (8.4\%) & 241 (12.9\%) & 337 (11.2\%) \\
\tabindent Yes & 1044 (91.6\%) & 1623 (87.1\%) & 2667 (88.8\%) \\ \bottomrule
\end{tabular}
\end{table}

The definition of the terms used in Figure \ref{fig:video-review-smry} are as follows.

\begin{itemize}
\item
  \textbf{Total clips}: The total number of video clips extracted for manual review.
\item
  \textbf{Bad clips}: Video clips for which all the analysis variables could not be extracted accurately (e.g., video quality issues, participants turned into a driveway right before a stop sign, the human reviewer could not properly assess the stopping behavior of a participant due to limited roadway visibility resulting from poor weather conditions).
\item
  \textbf{Good clips}: Video clips for which all the analysis variables could be extracted accurately.
\item
  \textbf{Discarded}: Driver trajectories with non-study drivers (i.e., a non-consented driver of the vehicle) or trajectories where the presence of vehicle or pedestrian affected the driver's approach speed at the intersection.
\item
  \textbf{Analyzable}: Driver trajectories where study participants were driving through stop intersections without any external interruption from other vehicles or pedestrians.
\end{itemize}

\section{Model structure}

\subsection{Modeling overview}
Stopping responses (unsafe vs. safe stopping) were modeled using mixed-effects logistic regression models (MELR) with by-participant and by-intersection random intercepts to account for driver- and intersection-level variation. The MELR models were fit using the ``lme4'' package \citep{bates_fitting_2015} version 1.1.25 for the R language and environment for statistical computing \citep{r_core_team_r_2021}.

We assessed 3 primary models to determine the impact of \textbf{worse sleep} (less sleep duration on average), \textbf{obesity} (greater BMI), and \textbf{worse glucose control} (GV: greater SD, CV, LGBI, HGBI) on the odds of unsafe stopping in drivers with diabetes. Sleep and obesity were assessed relative to drivers with and without diabetes. Glucose control was assessed in drivers with diabetes.

\subsection{Modeling variables}

\subsubsection{Outcome variable}
The three levels of the stopping responses (full stop, rolling stop, and no stop) exhibited by a participant driver at a stop sign were collapsed into two levels: unsafe stop (rolling stop, no stop) and safe stop (full stop). This was done to improve power since the ``no stop'' intersection encounters were very infrequent.

\subsubsection{Fixed effects}
\textbf{Sleep variables}: The average sleep duration (averaged across the entire study period) for each participant was centered and scaled (i.e., standardized) and was entered in the model as a continuous variable.

\textbf{Obesity variables}: The BMI measurements for participants were standardized and entered in the model as a continuous variable.

\begin{figure}[!htbp]
\centering
\includegraphics{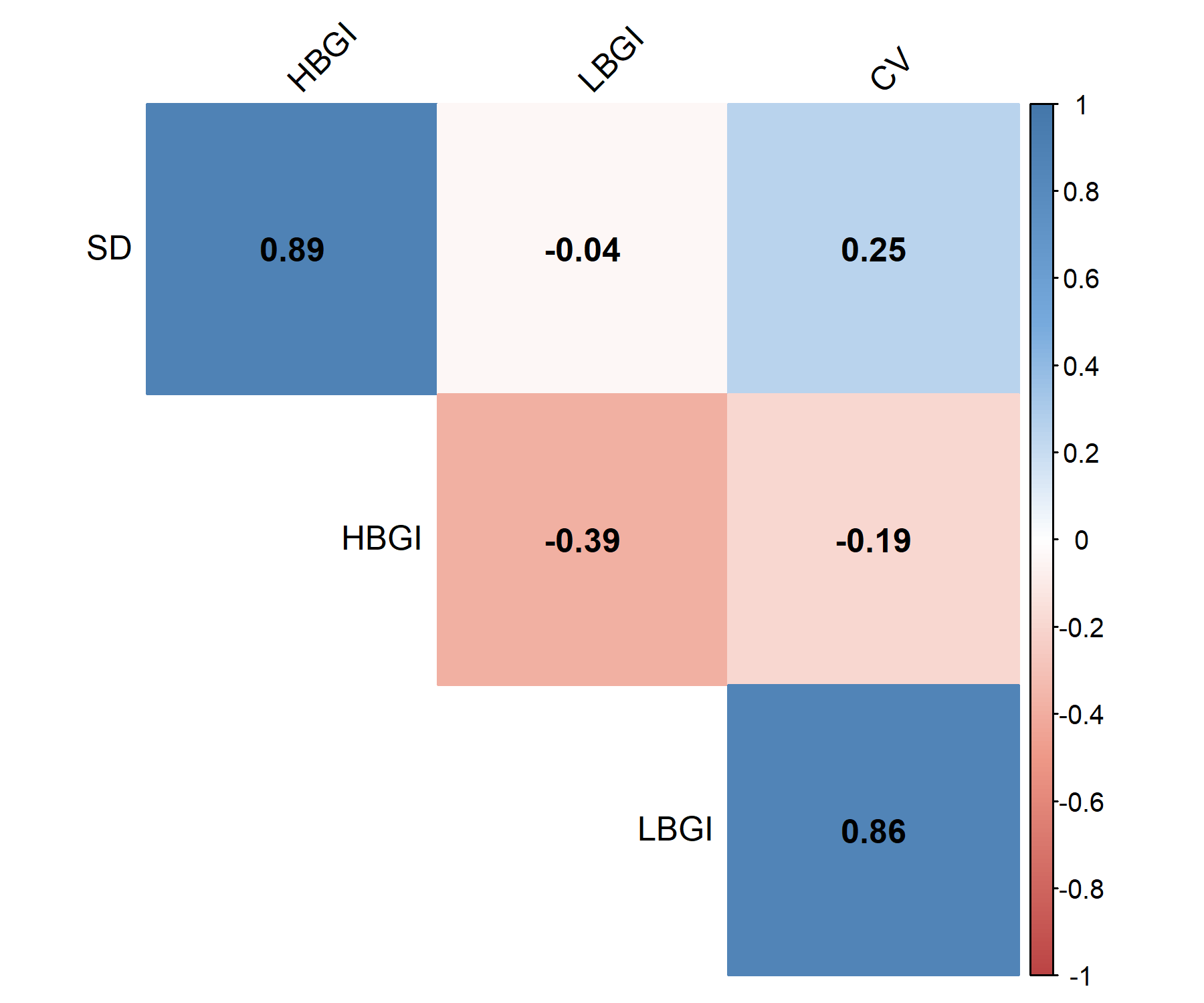}
\caption{Correlogram of GV variables}
\label{fig:gv-corrplot}
\end{figure}

\textbf{GV variables}: The four GV variables--SD, CV, HBGI, LBGI--were standardized and entered in the model as continuous variables. Additionally, correlations of the GV variables were also assessed (Figure \ref{fig:gv-corrplot}) to determine the model structure. Table \ref{tab:model-vars-descr-stats} provides the descriptive statistics for the fixed effects.

\begin{table}[!htbp]
\centering
\caption{Descriptive statistics table for sleep, obesity, and glucose control variables}
\label{tab:model-vars-descr-stats}
\begin{tabular}{@{}lccc@{}}
\toprule
 & \multicolumn{1}{c}{\textbf{T1DM (\textit{N} = 18)}} & \multicolumn{1}{c}{\textbf{Control (\textit{N} = 14)}} & \multicolumn{1}{c}{\textbf{Total (\textit{N} = 32)}} \\ \midrule
\textbf{Avg. sleep duration (hours)} &  &  &  \\
\tabindent N-Miss & 1 & 1 & 2 \\
\tabindent Mean (SD) & 6.80 (1.93) & 7.36 (0.85) & 7.05 (1.56) \\
\tabindent Range & 1.43--9.04 & 5.69--8.58 & 1.43--9.04 \\
\textbf{Body mass index (kg/m\textsuperscript{2})} &  &  &  \\
\tabindent Mean (SD) & 28.30 (5.21) & 27.88 (8.23) & 28.12 (6.58) \\
\tabindent Range & 20.80--38.97 & 20.77--49.78 & 20.77--49.78 \\
\textbf{LBGI} &  &  &  \\
\tabindent N-Miss & 0 & 14 & 14 \\
\tabindent Mean (SD) & 4.82 (2.70) & NA & 4.82 (2.70) \\
\tabindent Range & 2.06--11.32 & NA & 2.06--11.32 \\
\textbf{SD} &  &  &  \\
\tabindent N-Miss & 0 & 14 & 14 \\
\tabindent Mean (SD) & 71.43 (12.84) & NA & 71.43 (12.84) \\
\tabindent Range & 54.84--99.34 & NA & 54.84--99.34 \\
\textbf{HBGI} &  &  &  \\
\tabindent N-Miss & 0 & 14 & 14 \\
\tabindent Mean (SD) & 13.97 (5.85) & NA & 13.97 (5.85) \\
\tabindent Range & 6.63--28.41 & NA & 6.63--28.41 \\
\textbf{CV} &  &  &  \\
\tabindent N-Miss & 0 & 14 & 14 \\
\tabindent Mean (SD) & 0.40 (0.06) & NA & 0.40 (0.06) \\
\tabindent Range & 0.35--0.59 & NA & 0.35--0.59 \\ \bottomrule
\end{tabular}
\end{table}

\subsection{Model building procedure}
We used the following sequence of steps to build each model:

\begin{itemize}
\item
  \textbf{Step 1: Define a full model}: We started the model fitting process by defining a full model that included all fixed effects of interest and by-participant and by-intersection random intercepts. Two-way interactions between fixed effects were also included depending on the type of the model (e.g., sleep, glucose control) considered.
\item
  \textbf{Step 2: Select optimal random effects structure}: The full model defined in Step 1 was fitted with and without the by-intersection random intercept for the same fixed effects and the model fits were compared using the analysis of variance (ANOVA) test. If the result of the ANOVA test was statistically significant, the random effects structure with the by-intersection random intercept was selected as optimal, otherwise the by-intersection random intercept was eliminated.
\item
  \textbf{Step 3: Perform outlier data assessment}: Subsequent to the selection of the optimal random effects in Step 2, Cook's distance values \citep{cook_detection_1977, cook_influential_1979} were calculated for each participant and each intersection (calculated only if the by-intersection random intercept was selected in Step 2) to identify outlier participant(s) and/or intersection(s). Cook's distance quantifies the amount of change in model coefficients if data from a participant or an intersection is omitted. A participant or an intersection was identified as an outlier if its Cook's distance value was >0.5 \citep{pardoe_applied_2012} or it visually stood out compared to the Cook's distance values of other participants or intersections.
\item
  \textbf{Step 4: Interpret final model output}: Finally, we described the model fit results of the final model defined with the optimal random effects structure identified in Step 2 and without data from the outlier participant(s) and/or intersection(s) identified in Step 3.
\end{itemize}

\section{Results}

\subsection{Model 1: Sleep}

\subsubsection{Step 1: Define a full model}
To test the sleep hypothesis (T1DM + control drivers), we started with defining a full model that included age, gender, participant type, average sleep duration, and the two-way interaction between average sleep duration and participant type as fixed effects, and by-participant and by-intersection intercepts as random effects.

\subsubsection{Step 2: Select optimal random effects structure}
We fit the sleep model defined in Step 1 with and without the by-intersection random intercept. The model with the by-intersection random intercept was selected (\(\chi\left(1\right) = 30.62, p = <.001\); Table \ref{tab:sleep-opt-reff}).

\begin{table}[!htbp]
\centering
\caption{Sleep model fit output with and without the by-intersection random intercept}
\label{tab:sleep-opt-reff}
\begin{tabular}{@{}lcccc@{}}
\toprule
 & \multicolumn{2}{c}{\textbf{\begin{tabular}[c]{@{}c@{}}Sleep model \\ (with intxn reff) \end{tabular}}} & \multicolumn{2}{c}{\textbf{\begin{tabular}[c]{@{}c@{}}Sleep model \\ (without intxn reff) \end{tabular}}} \\ 
\textit{Predictors} & \multicolumn{1}{c}{\textit{OR}} & \textit{p} & \multicolumn{1}{c}{\textit{OR}} & \textit{p} \\ \midrule
Intercept & \multicolumn{1}{c}{\begin{tabular}[c]{@{}c@{}}3.33 \\ (0.85--13.08) \end{tabular}} & 0.086 & \multicolumn{1}{c}{\begin{tabular}[c]{@{}c@{}}2.81 \\ (0.78--10.08) \end{tabular}} & 0.113 \\
Age & \multicolumn{1}{c}{\begin{tabular}[c]{@{}c@{}}0.43 \\ (0.22--0.83) \end{tabular}} & \textbf{0.012} & \multicolumn{1}{c}{\begin{tabular}[c]{@{}c@{}}0.51 \\ (0.28--0.94) \end{tabular}} & \textbf{0.030} \\
Gender: Male & \multicolumn{1}{c}{\begin{tabular}[c]{@{}c@{}}2.48 \\ (0.63--9.80) \end{tabular}} & 0.194 & \multicolumn{1}{c}{\begin{tabular}[c]{@{}c@{}}2.26 \\ (0.63--8.07) \end{tabular}} & 0.209 \\
Participant type: T1DM & \multicolumn{1}{c}{\begin{tabular}[c]{@{}c@{}}1.57 \\ (0.34--7.31) \end{tabular}} & 0.564 & \multicolumn{1}{c}{\begin{tabular}[c]{@{}c@{}}1.57 \\ (0.37--6.60) \end{tabular}} & 0.537 \\
Avg. sleep duration & \multicolumn{1}{c}{\begin{tabular}[c]{@{}c@{}}3.60 \\ (0.22--57.64) \end{tabular}} & 0.366 & \multicolumn{1}{c}{\begin{tabular}[c]{@{}c@{}}3.08 \\ (0.23--40.51) \end{tabular}} & 0.393 \\
\begin{tabular}[c]{@{}l@{}}Participant type: T1DM x \\ Avg. sleep duration \end{tabular} & \multicolumn{1}{c}{\begin{tabular}[c]{@{}c@{}}0.27 \\ (0.01--5.19) \end{tabular}} & 0.383 & \multicolumn{1}{c}{\begin{tabular}[c]{@{}c@{}}0.32 \\ (0.02--4.96) \end{tabular}} & 0.413 \\
\textbf{Random Effects} & \multicolumn{4}{l}{} \\
\(\sigma^2\) & \multicolumn{2}{l}{3.29} & \multicolumn{2}{l}{3.29} \\
\(\tau_{00}\) & \multicolumn{2}{l}{1.11\textsubscript{intxn}} & \multicolumn{2}{l}{1.85\textsubscript{subj}} \\
 & \multicolumn{2}{l}{2.00\textsubscript{subj}} & \multicolumn{2}{l}{} \\
ICC & \multicolumn{2}{l}{0.49} & \multicolumn{2}{l}{0.36} \\
N & \multicolumn{2}{l}{28\textsubscript{subj}} & \multicolumn{2}{l}{28\textsubscript{subj}} \\
 & \multicolumn{2}{l}{473\textsubscript{intxn}} & \multicolumn{2}{l}{} \\
Observations & \multicolumn{2}{l}{1519} & \multicolumn{2}{l}{1519} \\
Marginal R\textsuperscript{2} / Conditional R\textsuperscript{2} & \multicolumn{2}{l}{0.147 / 0.562} & \multicolumn{2}{l}{0.128 / 0.441} \\ \bottomrule
\end{tabular}
\end{table}

The definition of different components of the random effects estimates shown in Table \ref{tab:sleep-opt-reff} are as follows: (a) \(\sigma^2\) is the within-group (or residual) variance. For MELR models, its value is fixed to \(\pi^{2}/3\); (b) \(\tau_{00}\) is the between-group random intercept variance; (c) ICC is the intraclass correlation coefficient; and (d) N is the number of random effect groups. Additionally, ``subj'', ``intxn'', and ``reff'' are shorthands for participant, intersection, and random effects, respectively.

\subsubsection{Step 3: Perform outlier data assessment}
\begin{figure}[!htbp]
\centering
\includegraphics[width = \textwidth]{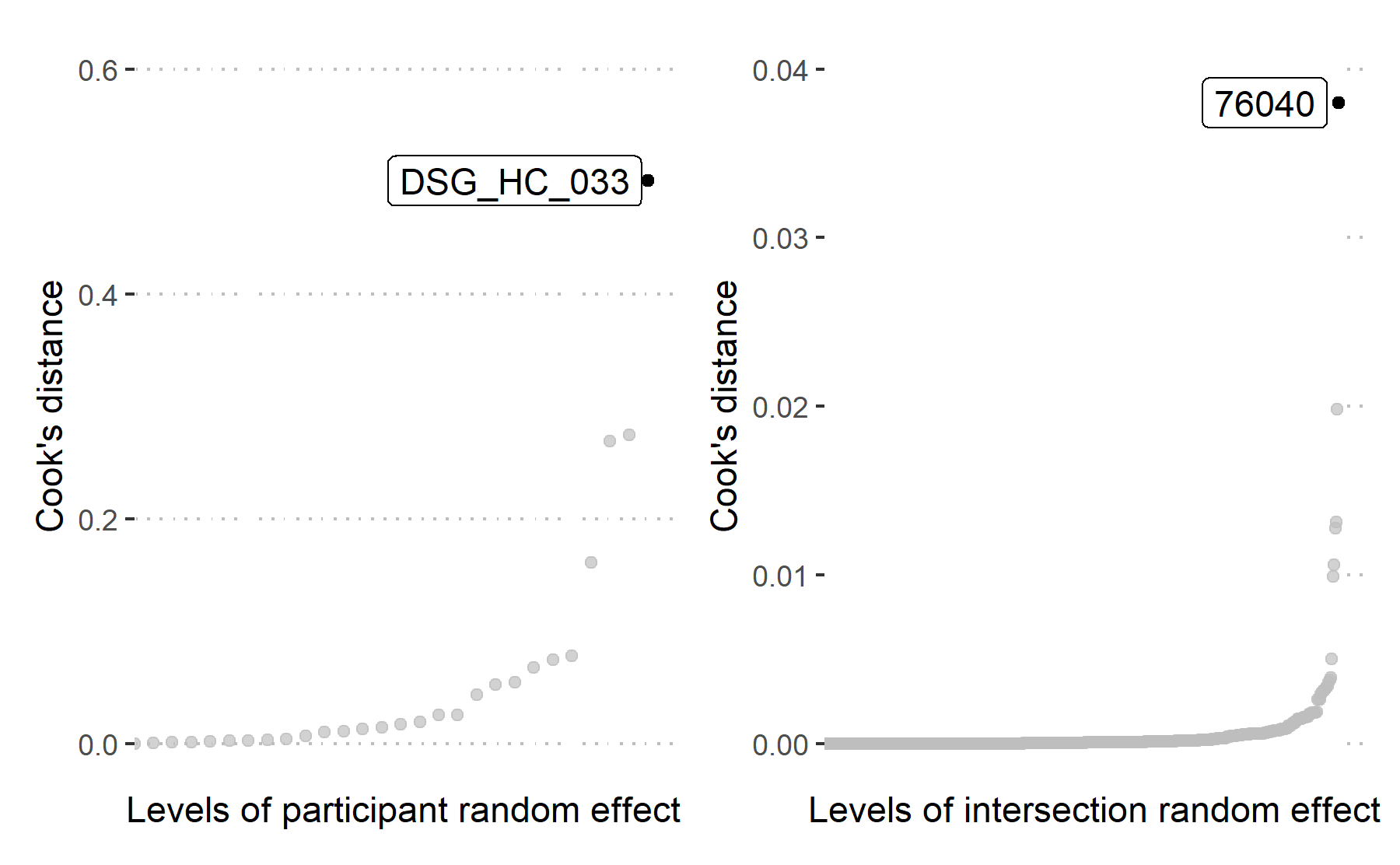}
\caption{Sleep model outlier assessments using Cook's distance}
\label{fig:sleep-cooksd-plot}
\end{figure}

Using the sleep model selected in Step 2, we performed outlier data assessments to check for outliers. Participant DSG\_HC\_033 (Cook's D = 0.50) and intersection ID 76040 were identified as outliers (Figure \ref{fig:sleep-cooksd-plot}) and the sleep model was refit by excluding the outliers (Table \ref{tab:sleep-outlier-assessment}).

\begin{table}[!htbp]
\centering
\caption{Sleep model fit output with and without outliers}
\label{tab:sleep-outlier-assessment}
\begin{tabular}{@{}lcccc@{}}
\toprule
 & \multicolumn{2}{c}{\textbf{\begin{tabular}[c]{@{}c@{}}Sleep model \\ (selected in Step 2) \end{tabular}}} & \multicolumn{2}{c}{\textbf{\begin{tabular}[c]{@{}c@{}}Sleep model \\ (without outliers) \end{tabular}}} \\ 
\textit{Predictors} & \multicolumn{1}{c}{\textit{OR}} & \textit{p} & \multicolumn{1}{c}{\textit{OR}} & \textit{p} \\ \midrule
Intercept & \multicolumn{1}{c}{\begin{tabular}[c]{@{}c@{}}3.33 \\ (0.85--13.08) \end{tabular}} & 0.086 & \multicolumn{1}{c}{\begin{tabular}[c]{@{}c@{}}9.24 \\ (2.25--37.95) \end{tabular}} & \textbf{0.002} \\
Age & \multicolumn{1}{c}{\begin{tabular}[c]{@{}c@{}}0.43 \\ (0.22--0.83) \end{tabular}} & \textbf{0.012} & \multicolumn{1}{c}{\begin{tabular}[c]{@{}c@{}}0.57 \\ (0.30--1.07) \end{tabular}} & 0.079 \\
Gender: Male & \multicolumn{1}{c}{\begin{tabular}[c]{@{}c@{}}2.48 \\ (0.63--9.80) \end{tabular}} & 0.194 & \multicolumn{1}{c}{\begin{tabular}[c]{@{}c@{}}2.47 \\ (0.72--8.45) \end{tabular}} & 0.148 \\
Participant type: T1DM & \multicolumn{1}{c}{\begin{tabular}[c]{@{}c@{}}1.57 \\ (0.34--7.31) \end{tabular}} & 0.564 & \multicolumn{1}{c}{\begin{tabular}[c]{@{}c@{}}0.58 \\ (0.13--2.71) \end{tabular}} & 0.491 \\
Avg. sleep duration & \multicolumn{1}{c}{\begin{tabular}[c]{@{}c@{}}3.60 \\ (0.22--57.64) \end{tabular}} & 0.366 & \multicolumn{1}{c}{\begin{tabular}[c]{@{}c@{}}0.70 \\ (0.04--11.03) \end{tabular}} & 0.801 \\
\begin{tabular}[c]{@{}l@{}}Participant type: T1DM x \\ Avg. sleep duration \end{tabular} & \multicolumn{1}{c}{\begin{tabular}[c]{@{}c@{}}0.27 \\ (0.01--5.19) \end{tabular}} & 0.383 & \multicolumn{1}{c}{\begin{tabular}[c]{@{}c@{}}1.37 \\ (0.07--26.37) \end{tabular}} & 0.834 \\
\textbf{Random Effects} & \multicolumn{4}{l}{} \\
\(\sigma^2\) & \multicolumn{2}{l}{3.29} & \multicolumn{2}{l}{3.29} \\
\(\tau_{00}\) & \multicolumn{2}{l}{1.11\textsubscript{intxn}} & \multicolumn{2}{l}{1.12\textsubscript{intxn}} \\
 & \multicolumn{2}{l}{2.00\textsubscript{subj}} & \multicolumn{2}{l}{1.50\textsubscript{subj}} \\
ICC & \multicolumn{2}{l}{0.49} & \multicolumn{2}{l}{0.44} \\
N & \multicolumn{2}{l}{28\textsubscript{subj}} & \multicolumn{2}{l}{27\textsubscript{subj}} \\
 & \multicolumn{2}{l}{473\textsubscript{intxn}} & \multicolumn{2}{l}{454\textsubscript{intxn}} \\
Observations & \multicolumn{2}{l}{1519} & \multicolumn{2}{l}{1429} \\
Marginal R\textsuperscript{2} / Conditional R\textsuperscript{2} & \multicolumn{2}{l}{0.147 / 0.562} & \multicolumn{2}{l}{0.094 / 0.496} \\ \bottomrule
\end{tabular}
\end{table}

\subsubsection{Step 4: Interpret model output}
The sleep model results presented in Table \ref{tab:sleep-outlier-assessment} indicates that omitting outliers did not change any model conclusions. Average sleep duration was non-significantly associated with the odds of making unsafe stops at stop intersections with or without the outliers.

\subsection{Model 2: Obesity}

\subsubsection{Step 1: Define a full model}
To test the obesity hypothesis (T1DM + control drivers), we defined a full model including age, gender, participant type, BMI, and the two-way interaction between BMI and participant type as fixed effects. By-participant and by-intersection intercepts were included as random effects.

\subsubsection{Step 2: Select optimal random effects structure}
We fit the obesity model defined in Step 1 with and without the by-intersection random intercept. The model with the by-intersection random intercept was selected (\(\chi\left(1\right) = 30.86, p = <.001\); Table \ref{tab:obesity-opt-reff}).

\begin{table}[!htbp]
\centering
\caption{Obesity model fit output with and without the by-intersection random intercept}
\label{tab:obesity-opt-reff}
\begin{tabular}{@{}lcccc@{}}
\toprule
 & \multicolumn{2}{c}{\textbf{\begin{tabular}[c]{@{}c@{}}Obesity model \\ (with intxn reff) \end{tabular}}} & \multicolumn{2}{c}{\textbf{\begin{tabular}[c]{@{}c@{}}Obesity model \\ (without intxn reff) \end{tabular}}} \\
\textit{Predictors} & \multicolumn{1}{c}{\textit{OR}} & \textit{p} & \multicolumn{1}{c}{\textit{OR}} & \textit{p} \\ \midrule
Intercept & \multicolumn{1}{c}{\begin{tabular}[c]{@{}c@{}}3.57 \\ (1.30--9.85) \end{tabular}} & \textbf{0.014} & \multicolumn{1}{c}{\begin{tabular}[c]{@{}c@{}}2.96 \\ (1.15--7.59) \end{tabular}} & 0.024 \\
Age & \multicolumn{1}{c}{\begin{tabular}[c]{@{}c@{}}0.62 \\ (0.34--1.14) \end{tabular}} & 0.123 & \multicolumn{1}{c}{\begin{tabular}[c]{@{}c@{}}0.71 \\ (0.41--1.24) \end{tabular}} & 0.230 \\
Gender: Male & \multicolumn{1}{c}{\begin{tabular}[c]{@{}c@{}}3.48 \\ (0.99--12.17) \end{tabular}} & 0.051 & \multicolumn{1}{c}{\begin{tabular}[c]{@{}c@{}}3.04 \\ (0.95--9.71) \end{tabular}} & 0.060 \\
Participant type: T1DM & \multicolumn{1}{c}{\begin{tabular}[c]{@{}c@{}}1.31 \\ (0.42--4.16) \end{tabular}} & 0.642 & \multicolumn{1}{c}{\begin{tabular}[c]{@{}c@{}}1.35 \\ (0.46--3.96) \end{tabular}} & 0.583 \\
BMI & \multicolumn{1}{c}{\begin{tabular}[c]{@{}c@{}}0.26 \\ (0.08--0.82) \end{tabular}} & \textbf{0.021} & \multicolumn{1}{c}{\begin{tabular}[c]{@{}c@{}}0.29 \\ (0.10--0.86) \end{tabular}} & \textbf{0.025} \\
\begin{tabular}[c]{@{}l@{}}Participant type: T1DM x \\ BMI \end{tabular} & \multicolumn{1}{c}{\begin{tabular}[c]{@{}c@{}}5.54 \\ (1.31--23.36) \end{tabular}} & \textbf{0.020} & \multicolumn{1}{c}{\begin{tabular}[c]{@{}c@{}}4.76 \\ (1.25--18.12) \end{tabular}} & \textbf{0.022} \\
\textbf{Random Effects} & \multicolumn{4}{l}{} \\
\(\sigma^2\) & \multicolumn{2}{l}{3.29} & \multicolumn{2}{l}{3.29} \\
\(\tau_{00}\) & \multicolumn{2}{l}{1.12\textsubscript{intxn}} & \multicolumn{2}{l}{1.57 \textsubscript{subj}} \\
 & \multicolumn{2}{l}{1.68\textsubscript{subj}} & \multicolumn{2}{l}{} \\
ICC & \multicolumn{2}{l}{0.46} & \multicolumn{2}{l}{0.32} \\
N & \multicolumn{2}{l}{28\textsubscript{subj}} & \multicolumn{2}{l}{28\textsubscript{subj}} \\
 & \multicolumn{2}{l}{473\textsubscript{intxn}} & \multicolumn{2}{l}{} \\
Observations & \multicolumn{2}{l}{1519} & \multicolumn{2}{l}{1519} \\
Marginal R\textsuperscript{2} / Conditional R\textsuperscript{2} & \multicolumn{2}{l}{0.211 / 0.574} & \multicolumn{2}{l}{0.196 / 0.456} \\ \bottomrule
\end{tabular}
\end{table}

\subsubsection{Step 3: Perform outlier data assessment}
\begin{figure}[!htbp]
\centering
\includegraphics[width = \textwidth]{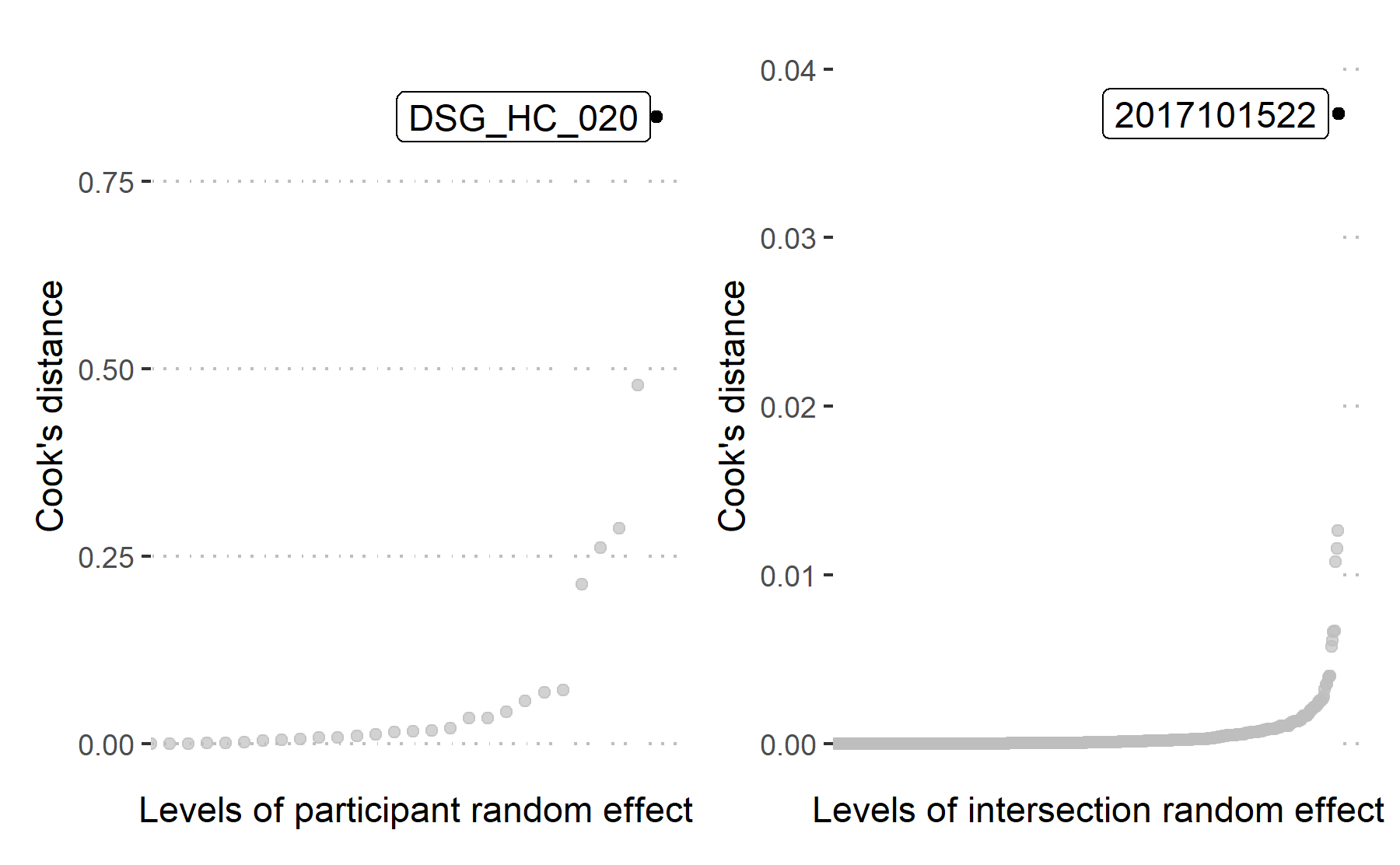}
\caption{Obesity model outlier assessments using Cook's distance}
\label{fig:obesity-cooksd-plot}
\end{figure}

Using the obesity model selected in Step 2, we performed outlier data assessment to check for outliers. Participant DSG\_HC\_020 (Cook's D = 0.84) and intersection ID 2017101522 were identified as outliers (Figure \ref{fig:obesity-cooksd-plot}) and the obesity model was refit by excluding the outliers (Table \ref{tab:obesity-outlier-assessment}).

\begin{table}[!htbp]
\centering
\caption{Obesity model fit output with and without outliers}
\label{tab:obesity-outlier-assessment}
\begin{tabular}{@{}lcccc@{}}
\toprule
 & \multicolumn{2}{c}{\textbf{\begin{tabular}[c]{@{}c@{}}Obesity model \\ (selected in Step 2) \end{tabular}}} & \multicolumn{2}{c}{\textbf{\begin{tabular}[c]{@{}c@{}}Obesity model \\ (without outliers) \end{tabular}}} \\ 
\textit{Predictors} & \multicolumn{1}{c}{\textit{OR}} & \textit{p} & \multicolumn{1}{c}{\textit{OR}} & \textit{p} \\ \midrule
Intercept & \multicolumn{1}{c}{\begin{tabular}[c]{@{}c@{}}3.57 \\ (1.30--9.85) \end{tabular}} & \textbf{0.014} & \multicolumn{1}{c}{\begin{tabular}[c]{@{}c@{}}1.78 \\ (0.67--4.69) \end{tabular}} & 0.246 \\
Age & \multicolumn{1}{c}{\begin{tabular}[c]{@{}c@{}}0.62 \\ (0.34--1.14) \end{tabular}} & 0.123 & \multicolumn{1}{c}{\begin{tabular}[c]{@{}c@{}}0.46 \\ (0.27--0.78) \end{tabular}} & \textbf{0.004} \\
Gender: Male & \multicolumn{1}{c}{\begin{tabular}[c]{@{}c@{}}3.48 \\ (0.99--12.17) \end{tabular}} & 0.051 & \multicolumn{1}{c}{\begin{tabular}[c]{@{}c@{}}3.88 \\ (1.30--11.54) \end{tabular}} & \textbf{0.015} \\
Participant type: T1DM & \multicolumn{1}{c}{\begin{tabular}[c]{@{}c@{}}1.31 \\ (0.42--4.16) \end{tabular}} & 0.642 & \multicolumn{1}{c}{\begin{tabular}[c]{@{}c@{}}2.54 \\ (0.88--7.35) \end{tabular}} & 0.086 \\
BMI & \multicolumn{1}{c}{\begin{tabular}[c]{@{}c@{}}0.26 \\ (0.08--0.82) \end{tabular}} & \textbf{0.021} & \multicolumn{1}{c}{\begin{tabular}[c]{@{}c@{}}0.12 \\ (0.04--0.36) \end{tabular}} & \textbf{<0.001} \\
\begin{tabular}[c]{@{}l@{}}Participant type: T1DM x \\ BMI \end{tabular} & \multicolumn{1}{c}{\begin{tabular}[c]{@{}c@{}}5.54 \\ (1.31--23.36) \end{tabular}} & \textbf{0.020} & \multicolumn{1}{c}{\begin{tabular}[c]{@{}c@{}}14.96 \\ (3.75--59.73) \end{tabular}} & \textbf{<0.001} \\
\textbf{Random Effects} & \multicolumn{4}{l}{} \\
\(\sigma^2\) & \multicolumn{2}{l}{3.29} & \multicolumn{2}{l}{3.29} \\
\(\tau_{00}\) & \multicolumn{2}{l}{1.12\textsubscript{intxn}} & \multicolumn{2}{l}{1.12\textsubscript{intxn}} \\
 & \multicolumn{2}{l}{1.68\textsubscript{subj}} & \multicolumn{2}{l}{1.14\textsubscript{subj}} \\
ICC & \multicolumn{2}{l}{0.46} & \multicolumn{2}{l}{0.41} \\
N & \multicolumn{2}{l}{28\textsubscript{subj}} & \multicolumn{2}{l}{27\textsubscript{subj}} \\
 & \multicolumn{2}{l}{473\textsubscript{intxn}} & \multicolumn{2}{l}{456\textsubscript{intxn}} \\
Observations & \multicolumn{2}{l}{1519} & \multicolumn{2}{l}{1480} \\
Marginal R\textsuperscript{2} / Conditional R\textsuperscript{2} & \multicolumn{2}{l}{0.211 / 0.574} & \multicolumn{2}{l}{0.363 / 0.622} \\ \bottomrule
\end{tabular}
\end{table}

\subsubsection{Step 4: Interpret model output}
The obesity model results presented in Table \ref{tab:obesity-outlier-assessment} indicate that removing outliers lowered the \emph{p}-values of fixed effects in the model and that all the fixed effects were statistically significant with outlier removal. Fixed effect direction did not change. Final results show a significant interaction between participant type and BMI, where a one standard deviation (\(\sim\)7 points) of increase in BMI increases unsafe stop odds by 14.96 for T1DM participants compared to controls.

\subsection{Model 3: Glucose control}

\subsubsection{Step 1: Define a full model}
To test the glucose control hypothesis (T1DM drivers only), we defined two full models, one for each pair of uncorrelated GV variables (LBGI and SD; HBGI and CV) (see Figure \ref{fig:gv-corrplot}). In addition to GV variables, the full models included age and gender as fixed effects, and by-participant and by-intersection intercepts as random effects.

\subsubsection{Step 2: Select optimal random effects structure}
We fit the glucose control models defined in Step 1 with and without the by-intersection random intercept. The models with the by-intersection random intercept were selected (LBGI + SD: \(\chi\left(1\right) = 4.03, p = .045\); HBGI + CV: \(\chi\left(1\right) = 4.16, p = .041\); Table \ref{tab:glucose-opt-reff}).

\begin{table}[!htbp]
\setlength{\tabcolsep}{2pt}
\small
\centering
\caption{Glucose control model fit output with and without the by-intersection random intercept}
\label{tab:glucose-opt-reff}
\begin{tabular}{@{}lcccccccc@{}}
\toprule
 & \multicolumn{2}{c}{\textbf{\begin{tabular}[c]{@{}c@{}}LBGI + SD \\ (with intxn reff) \end{tabular}}} & \multicolumn{2}{c}{\textbf{\begin{tabular}[c]{@{}c@{}}LBGI + SD \\ (without intxn reff) \end{tabular}}} & \multicolumn{2}{c}{\textbf{\begin{tabular}[c]{@{}c@{}}HBGI + CV \\ (with intxn reff) \end{tabular}}} & \multicolumn{2}{c}{\textbf{\begin{tabular}[c]{@{}c@{}}HBGI + CV \\ (without intxn reff) \end{tabular}}} \\ 
\textit{Predictors} & \textit{OR} & \textit{p} & \textit{OR} & \textit{p} & \textit{OR} & \textit{p} & \textit{OR} & \textit{p} \\ \midrule
Intercept & \begin{tabular}[c]{@{}c@{}}3.04 \\ (1.51--6.12) \end{tabular} & \textbf{0.002} & \begin{tabular}[c]{@{}c@{}}2.76 \\ (1.43--5.34) \end{tabular} & \textbf{0.003} & \begin{tabular}[c]{@{}c@{}}3.09 \\ (1.53--6.24) \end{tabular} & \textbf{0.002} & \begin{tabular}[c]{@{}c@{}}2.81 \\ (1.44--5.48) \end{tabular} & \textbf{0.002} \\
Age & \begin{tabular}[c]{@{}c@{}}0.46 \\ (0.25--0.85) \end{tabular} & \textbf{0.013} & \begin{tabular}[c]{@{}c@{}}0.51 \\ (0.29--0.90) \end{tabular} & \textbf{0.021} & \begin{tabular}[c]{@{}c@{}}0.48 \\ (0.27--0.87) \end{tabular} & \textbf{0.015} & \begin{tabular}[c]{@{}c@{}}0.54 \\ (0.31--0.94) \end{tabular} & \textbf{0.029} \\
Gender: Male & \begin{tabular}[c]{@{}c@{}}4.46 \\ (1.36--14.58) \end{tabular} & \textbf{0.013} & \begin{tabular}[c]{@{}c@{}}4.11 \\ (1.31--12.89) \end{tabular} & \textbf{0.015} & \begin{tabular}[c]{@{}c@{}}4.46 \\ (1.34--14.86) \end{tabular} & \textbf{0.015} & \begin{tabular}[c]{@{}c@{}}4.06 \\ (1.27--12.98) \end{tabular} & \textbf{0.018} \\
LBGI & \begin{tabular}[c]{@{}c@{}}1.19 \\ (0.67--2.11) \end{tabular} & 0.562 & \begin{tabular}[c]{@{}c@{}}1.22 \\ (0.70--2.14) \end{tabular} & 0.483 &  &  &  &  \\
SD & \begin{tabular}[c]{@{}c@{}}1.10 \\ (0.62--1.96) \end{tabular} & 0.747 & \begin{tabular}[c]{@{}c@{}}1.03 \\ (0.59--1.79) \end{tabular} & 0.914 &  &  &  &  \\
HBGI &  &  &  &  & \begin{tabular}[c]{@{}c@{}}1.11 \\ (0.62--1.99) \end{tabular} & 0.728 & \begin{tabular}[c]{@{}c@{}}1.03 \\ (0.59--1.79) \end{tabular} & 0.931 \\
CV &  &  &  &  & \begin{tabular}[c]{@{}c@{}}1.06 \\ (0.60--1.84) \end{tabular} & 0.850 & \begin{tabular}[c]{@{}c@{}}1.07 \\ (0.62--1.85) \end{tabular} & 0.808 \\
\textbf{Random Effects} & \multicolumn{8}{l}{} \\
\(\sigma^2\) & \multicolumn{2}{l}{3.29} & \multicolumn{2}{l}{3.29} & \multicolumn{2}{l}{3.29} & \multicolumn{2}{l}{3.29} \\
\(\tau_{00}\) & \multicolumn{2}{l}{0.70\textsubscript{intxn}} & \multicolumn{2}{l}{0.92\textsubscript{subj}} & \multicolumn{2}{l}{0.71\textsubscript{intxn}} & \multicolumn{2}{l}{0.95\textsubscript{subj}} \\
 & \multicolumn{2}{l}{0.89\textsubscript{subj}} & \multicolumn{2}{l}{} & \multicolumn{2}{l}{0.91\textsubscript{subj}} & \multicolumn{2}{l}{} \\
ICC & \multicolumn{2}{l}{0.33} & \multicolumn{2}{l}{0.22} & \multicolumn{2}{l}{0.33} & \multicolumn{2}{l}{0.22} \\
N & \multicolumn{2}{l}{18\textsubscript{subj}} & \multicolumn{2}{l}{18\textsubscript{subj}} & \multicolumn{2}{l}{18\textsubscript{subj}} & \multicolumn{2}{l}{18\textsubscript{subj}} \\
 & \multicolumn{2}{l}{243\textsubscript{intxn}} & \multicolumn{2}{l}{} & \multicolumn{2}{l}{243 intxn} & \multicolumn{2}{l}{} \\
Observations & \multicolumn{2}{l}{561} & \multicolumn{2}{l}{561} & \multicolumn{2}{l}{561} & \multicolumn{2}{l}{561} \\
\makecell[tl]{Marginal R\textsuperscript{2} / \\ Conditional R\textsuperscript{2}} & \multicolumn{2}{l}{0.180 / 0.447} & \multicolumn{2}{l}{0.166 / 0.349} & \multicolumn{2}{l}{0.189 / 0.456} & \multicolumn{2}{l}{0.173 / 0.359} \\ \bottomrule
\end{tabular}
\end{table}

\subsubsection{Step 3: Perform outlier data assessment}
\begin{figure}[!htbp]
\centering
\includegraphics[width = \textwidth]{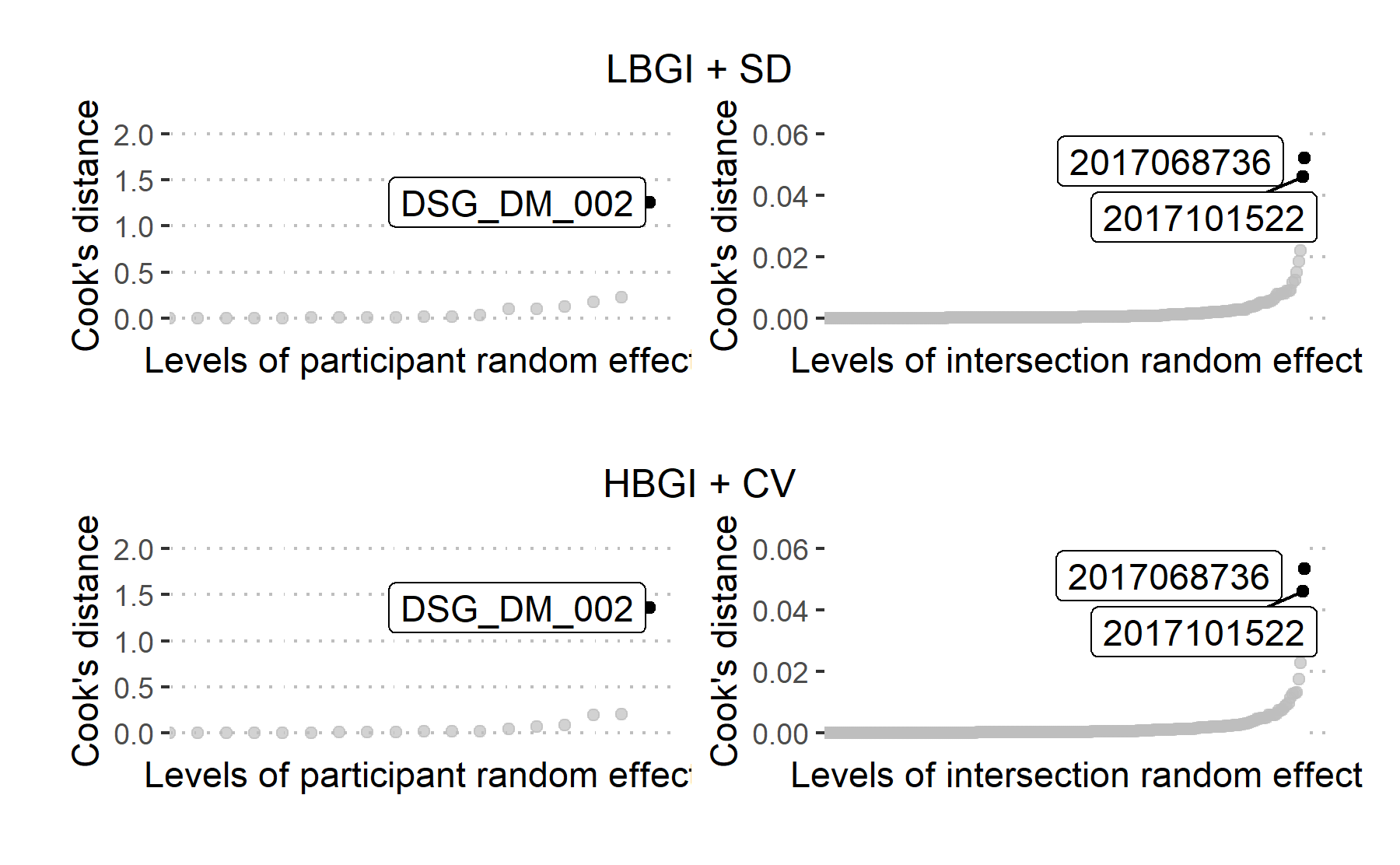}
\caption{Glucose control model outlier assessments using Cook's distance}
\label{fig:glucose-cooksd-plot}
\end{figure}

Using the glucose control models selected in Step 2, we performed outlier data assessment to check for outliers. Participant DSG\_DM\_002 [Cook's D = 1.26 (LBGI + SD model), 1.36 (HBGI + CV model)] and two intersections (IDs: 2017068736 and 2017101522) were identified as outliers (Figure \ref{fig:glucose-cooksd-plot}) and the glucose control models were refit by excluding the outliers (Table \ref{tab:glucose-outlier-assessment}).

\begin{table}[!htbp]
\setlength{\tabcolsep}{2pt}
\small
\centering
\caption{Glucose control fit output with and without outliers}
\label{tab:glucose-outlier-assessment}
\begin{tabular}{@{}lcccccccc@{}}
\toprule
 & \multicolumn{2}{c}{\textbf{\begin{tabular}[c]{@{}c@{}}LBGI + SD \\ (from Step 2) \end{tabular}}} & \multicolumn{2}{c}{\textbf{\begin{tabular}[c]{@{}c@{}}LBGI + SD \\    (without outliers) \end{tabular}}} & \multicolumn{2}{c}{\textbf{\begin{tabular}[c]{@{}c@{}}HBGI + CV \\ (from Step 2) \end{tabular}}} & \multicolumn{2}{c}{\textbf{\begin{tabular}[c]{@{}c@{}}HBGI + CV \\ (without outliers) \end{tabular}}} \\ 
\textit{Predictors} & \textit{OR} & \textit{p} & \textit{OR} & \textit{p} & \textit{OR} & \textit{p} & \textit{OR} & \textit{p} \\ \midrule
Intercept & \begin{tabular}[c]{@{}c@{}}3.04 \\ (1.51--6.12) \end{tabular} & \textbf{0.002} & \begin{tabular}[c]{@{}c@{}}5.48 \\ (3.18--9.44) \end{tabular} & \textbf{<0.001} & \begin{tabular}[c]{@{}c@{}}3.09 \\ (1.53--6.24) \end{tabular} & \textbf{0.002} & \begin{tabular}[c]{@{}c@{}}5.66 \\ (3.29--9.75) \end{tabular} & \textbf{<0.001} \\
Age & \begin{tabular}[c]{@{}c@{}}0.46 \\ (0.25--0.85) \end{tabular} & \textbf{0.013} & \begin{tabular}[c]{@{}c@{}}0.52 \\ (0.34--0.79) \end{tabular} & \textbf{0.002} & \begin{tabular}[c]{@{}c@{}}0.48 \\ (0.27--0.87) \end{tabular} & \textbf{0.015} & \begin{tabular}[c]{@{}c@{}}0.52 \\ (0.35--0.77) \end{tabular} & \textbf{0.001} \\
Gender: Male & \begin{tabular}[c]{@{}c@{}}4.46 \\ (1.36--14.58) \end{tabular} & \textbf{0.013} & \begin{tabular}[c]{@{}c@{}}1.84 \\ (0.84--4.06) \end{tabular} & 0.130 & \begin{tabular}[c]{@{}c@{}}4.46 \\ (1.34--14.86) \end{tabular} & \textbf{0.015} & \begin{tabular}[c]{@{}c@{}}1.78 \\ (0.82--3.88) \end{tabular} & 0.148 \\
LBGI & \begin{tabular}[c]{@{}c@{}}1.19 \\ (0.67--2.11) \end{tabular} & 0.562 & \begin{tabular}[c]{@{}c@{}}0.95 \\ (0.67--1.34) \end{tabular} & 0.762 &  &  &  &  \\
SD & \begin{tabular}[c]{@{}c@{}}1.10 \\ (0.62--1.96) \end{tabular} & 0.747 & \begin{tabular}[c]{@{}c@{}}0.82 \\ (0.55--1.23) \end{tabular} & 0.347 &  &  &  &  \\
HBGI &  &  &  &  & \begin{tabular}[c]{@{}c@{}}1.11 \\ (0.62--1.99) \end{tabular} & 0.728 & \begin{tabular}[c]{@{}c@{}}0.83 \\ (0.56--1.23) \end{tabular} & 0.361 \\
CV &  &  &  &  & \begin{tabular}[c]{@{}c@{}}1.06 \\ (0.60--1.84) \end{tabular} & 0.850 & \begin{tabular}[c]{@{}c@{}}0.82 \\ (0.59--1.15) \end{tabular} & 0.244 \\
\textbf{Random Effects} & \multicolumn{8}{l}{} \\
\(\sigma^2\) & \multicolumn{2}{l}{3.29} & \multicolumn{2}{l}{3.29} & \multicolumn{2}{l}{3.29} & \multicolumn{2}{l}{3.29} \\
\(\tau_{00}\) & \multicolumn{2}{l}{0.70\textsubscript{intxn}} & \multicolumn{2}{l}{0.70 intxn} & \multicolumn{2}{l}{0.71\textsubscript{intxn}} & \multicolumn{2}{l}{0.76\textsubscript{intxn}} \\
 & \multicolumn{2}{l}{0.89\textsubscript{subj}} & \multicolumn{2}{l}{0.11\textsubscript{subj}} & \multicolumn{2}{l}{0.91\textsubscript{subj}} & \multicolumn{2}{l}{0.07\textsubscript{subj}} \\
ICC & \multicolumn{2}{l}{0.33} & \multicolumn{2}{l}{0.20} & \multicolumn{2}{l}{0.33} & \multicolumn{2}{l}{0.20} \\
N & \multicolumn{2}{l}{18\textsubscript{subj}} & \multicolumn{2}{l}{17\textsubscript{subj}} & \multicolumn{2}{l}{18\textsubscript{subj}} & \multicolumn{2}{l}{17\textsubscript{subj}} \\
 & \multicolumn{2}{l}{243\textsubscript{intxn}} & \multicolumn{2}{l}{225\textsubscript{intxn}} & \multicolumn{2}{l}{243\textsubscript{intxn}} & \multicolumn{2}{l}{225\textsubscript{intxn}} \\
Observations & \multicolumn{2}{l}{561} & \multicolumn{2}{l}{522} & \multicolumn{2}{l}{561} & \multicolumn{2}{l}{522} \\
\makecell[tl]{Marginal R\textsuperscript{2} / \\ Conditional R\textsuperscript{2}} & \multicolumn{2}{l}{0.180 / 0.447} & \multicolumn{2}{l}{0.108 / 0.285} & \multicolumn{2}{l}{0.189 / 0.456} & \multicolumn{2}{l}{0.112 / 0.291} \\ \bottomrule
\end{tabular}
\end{table}

\subsubsection{Step 4: Interpret model output}
The glucose model results presented in Table \ref{tab:glucose-outlier-assessment} indicate that omitting outliers did not change model conclusions and GV variables were non-significantly associated with the odds of making unsafe stops at stop intersections with or without the outliers.

\subsection{Control variable results}
Across the three models, the association between the control variables (age, gender) and the unsafe stopping behavior was similar. Table \ref{tab:control-vars-fit-smry} provides a summary of the model fit results for the control variables.

\begin{table}[!htbp]
\centering
\caption{Summary of model fit results for control variables}
\label{tab:control-vars-fit-smry}
\begin{tabular}{@{}lcccc@{}}
\toprule
\multicolumn{1}{c}{\textbf{Model}} & \multicolumn{2}{c}{\textbf{Age}} & \multicolumn{2}{c}{\textbf{Gender: Male}} \\ 
 & \textit{OR} & \textit{p-value} & \textit{OR} & \textit{p-value} \\ \midrule
Sleep & 0.57 & 0.079 & 2.47 & 0.148 \\
Obesity & 0.46 & \textbf{0.004} & 3.88 & \textbf{0.015} \\
Glucose: LBGI + SD & 0.52 & \textbf{0.002} & 1.84 & 0.130 \\
Glucose: HBGI + CV & 0.52 & \textbf{0.001} & 1.78 & 0.148 \\ \bottomrule
\end{tabular}
\end{table}

Based on the results presented in Table \ref{tab:control-vars-fit-smry}, age decreases the odds of an unsafe stop in all models, with drivers who were one standard deviation older (\(\sim\)10 years) were \(\sim\)50\% less likely to make an unsafe stop. Gender had little effect on unsafe stopping, with the exception of drivers who are male and more obese. Male drivers with more obesity (\(\sim\)7 BMI higher) had a 3.88 greater odds of unsafe stopping than similar female drivers.

\section{Conclusion}
This study provides novel insight into how chronic, at-risk physiology and health impact unsafe driver behavior in type-1 diabetes. A key finding is that greater BMI in T1DM drivers increases unsafe stopping frequency while sleep and chronic glucose impairments do not. In line with our previous research showing that acute glucose impairments affect driver risk \citep{chakraborty_quantifying_2019, merickel_driving_2019}, this suggests that obesity and acute glucose impairment are primary targets for clinical intervention in diabetes care. This finding reinforces FMCSA recommendations \citep{federal_motor_carrier_safety_administration_medical_2020} to evaluate obesity and associated co-morbid health conditions, like obstructive sleep apnea, when considering driver risk. In the context of clinical care, this stresses the need for doctors to consider obesity and weight control, along with acute glucose dysfunction, in T1DM patients when counseling and educating patients on real-world safety. While this study did not find that sleep dysfunction affected driver risk, this may be due to overall infrequent abnormal sleep in the population studied, suggesting need to assess sleep dysfunction in a broader population of diabetes drivers.

Although this paper presents some promising findings, there are admittedly targets for future research. A limitation of this paper is the small sample size of 32 participants and the short study duration of 4-weeks. More varied participant samples, along with longer observations of driver physiology and sleep may provide further understanding of how chronic driver impairments affect risk. Future research may also focus on the development of automated methods capable of performing facial behavior analyses on driver cabin videos to extract information on head pose and eye-gaze as drivers approach an intersection. Another focus area could be the development of automated methods capable of encoding interactions between drivers and surrounding traffic and capturing the stopping behavior of drivers with high degree of accuracy. The availability of such methods for video data processing will greatly cut down the human labor and time needed for the preparation of analysis-ready data sets.

Overall, the results show that wearable and in-vehicle sensor technology can successfully be used to measure and assess at-risk, real-world driver behavior. A particularly promising application of this study is developing improved driver assistance technology capable of sensing, indexing, and responding to driver health and intervening to prevent risk. This technology holds promise for supporting mobility and independence in patients with diabetes and other diseases, while minimizing disease burden. With technology advancements, vehicles may be able to collect, parse, and deliver data to healthcare providers for continuous patient health and risk monitoring, along with real-time risk intervention and driver feedback.

\section{Acknowledgements}
We gratefully acknowledge the Toyota Collaborative Safety Research Center for funding this study and the Mind \& Brain Health Labs at UNMC's Department of Neurological Sciences for leading study operations. We extend our thanks particularly to Dr. Andjela Drincic for her invaluable clinical work and guidance on all aspects of this study.

\Urlmuskip=0mu plus 1mu\relax
\bibliographystyle{elsarticle-harv}
\clearpage
\bibliography{references}

\end{document}